\documentclass[journal,twoside,web]{ieeecolor}
\usepackage{generic}
\usepackage{cite}
\usepackage{amsmath,amssymb,amsfonts}
\usepackage{algorithmic}
\usepackage{graphicx}
\usepackage{multirow}
\usepackage{booktabs}
\usepackage{textcomp}
\def\BibTeX{{\rm B\kern-.05em{\sc i\kern-.025em b}\kern-.08em
		T\kern-.1667em\lower.7ex\hbox{E}\kern-.125emX}}
\begin{document}
	\title{EEG-based Epileptic Prediction via a Two-stage Channel-aware Set Transformer Network}
	
	\author{Ruifeng Zheng,
		Cong Chen,
		Shuang Wang,
		Yiming Liu,
		Lin You,
		Jindong Lu, 
		Ruizhe Zhu,
		Guodao Zhang,
		and Kejie Huang
		\thanks{Corresponding author: Kejie Huang (e-mail: huangkejie@zju.edu.cn)}
	}
	
	\maketitle
	
	\begin{abstract}
		Epilepsy is a chronic, noncommunicable brain disorder, and sudden seizure onsets can significantly impact patients' quality of life and health. However, wearable seizure-predicting devices are still limited, partly due to the bulky size of EEG-collecting devices. To relieve the problem, we proposed a novel two-stage channel-aware Set Transformer Network that could perform seizure prediction with fewer EEG channel sensors. We also tested a seizure-independent division method which could prevent the adjacency of training and test data. Experiments were performed on the CHB-MIT dataset which includes 22 patients with 88 merged seizures. The mean sensitivity before channel selection was 76.4\% with a false predicting rate (FPR) of 0.09/hour. After channel selection, dominant channels emerged in 20 out of 22 patients; the average number of channels was reduced to 2.8 from 18; and the mean sensitivity rose to 80.1\% with an FPR of 0.11/hour. Furthermore, experimental results on the seizure-independent division supported our assertion that a more rigorous seizure-independent division should be used for patients with abundant EEG recordings.
		
	\end{abstract}
	
	
	\section{Introduction}
	Epilepsy is a chronic noncommunicable brain disease that affects around fifty million people globally. \cite{world2019epilepsy} Approximately one-third of epilepsy patients grapple with intractable epilepsy\cite{thurman2011standards}. The sudden onset of epilepsy contributes to anxiety in patients' daily life and may result in accidents that threaten their health and even their lives. Effective seizure prediction can provide early warnings to patients, alleviating the risks posed by seizures. It can also prompt wearable medical devices to initiate clinical interventions \cite{chen2016transcranial,yan2022seizure} to suppress imminent seizure onset. Electroencephalogram (EEG) has been widely adopted for epilepsy diagnosis due to its non-invasive nature. 
	
	However, the commercial application of wearable EEG collection devices is limited. This can be partly attributed to their discomfort and inconvenience for all-day wear, especially considering their often bulky size. If the number of electrode channels could be reduced, the devices would be implemented with fewer sensors, leading to a smaller size, reduced power consumption, and lower cost. To reduce the number of sensors, we proposed a novel two-stage channel-aware Set Transformer Network which can identify dominant electrode channels for seizure prediction. Specifically, Set Transformer \cite{vaswani2017attention, lee2019set}, which features permutation invariant and has a lower computational overhead compared to standard Transformers, was employed to organize EEG features. The reason for the selection is our observation that certain segments of the EEG signal and certain electrode channels hold greater predictive importance than others, and the specific order in which they are arranged is less significant. The input features were merged temporally by a Set Transformer in the first stage of our network. In the second stage, we developed a channel-aware Set Transformer to process the temporally-merged features obtained from the first stage. This channel-aware Transformer incorporated features from all channels and conducted patient-specific channel selection when the prediction results for a given patient were heavily influenced by a few specific channels.
	
	
	Furthermore, we tested a more rigorous method for dividing EEG data. In many previous studies, all interictal EEG signals were concatenated together and then divided evenly (this dividing method will be referred to as ``even division" in this article for brevity). However, the even division usually fragments an integrated interictal sequence into pieces that can appear in both training and test data. Therefore, we tested a seizure-independent division which is closer to clinical practice. Among this division, an interictal EEG sequence before a specific seizure acts as either training or test data, avoiding the adjacency of training and test data.

	Additionally, many prior studies chose to leave out four-hour EEG signals before or after seizures to constitute the interictal data, likely leading to alarms being triggered four hours before a seizure. To mitigate this issue, our study only excluded one-hour data before or after seizures to form the interictal data.
	
	EEG data of patients from CHB-MIT database \cite{shoeb2009application} were used in this study. Twenty-two patients were included, and 18 channels shared by all patients were selected as input channels. Several band power features of EEG signals were chosen as the key indicators for prediction. The seizure prediction horizon (SPH) and seizure occurrence period (SOP) were set to three and 30 minutes, respectively. Adjacent seizures with an interval of less than one hour were merged, and seizures with preictal lengths significantly less than 30 minutes were omitted. The even division was conducted on 22 patients with 88 merged seizures, and the seizure-independent division was conducted on seven patients with 26 qualified seizures.
	
	Experiment results on the even division demonstrate that our two-stage channel-aware Set Transformer has achieved good performance in epileptic prediction. Before channel selection, the mean sensitivity was 76.4\% with a false predicting rate of 0.09 per hour. After introducing channel selection, dominant channels emerged in 20 out of 22 patients. A mean sensitivity of 80.1\% with an FPR of 0.11 per hour were achieved after channel selection. The average number of channels was reduced to 2.8 after the selection, such that EEG collecting devices can be implemented with smaller sizes and less power consumption, thus increasing the popularity among epilepsy patients. In addition, experiment results on the seizure-independent division support our suggestion that the seizure-independent division should be conducted for patients with abundant EEG recordings.
	
	In summary, our contributions are as follows:
	\begin{itemize}
		\item The bulky size of EEG collecting sensors hinders the popularity of wearable seizure predicting devices. To relieve the problem, we proposed a novel two-stage channel-aware Set Transformer Network which could perform seizure prediction with much fewer EEG collecting sensors.
		\item We employed low-complexity input features, and our solution took about 33.5 ms to handle an input EEG signal segment arriving every second, allowing for real-time monitoring and rapid response. 
		\item Our method achieves a mean sensitivity of 80.1\% with 0.11 false alarms per hour on patients in the CHB-MIT dataset on the even division, and the average number of electrode channels reduces to 2.8 from 18 after selection.
	\end{itemize}
	
	The paper is structured as follows: Section II introduces relevant studies about epileptic EEG signal analyses; Section III details our data preprocessing, the definition of the seizure-independent division, and the structure of the two-stage channel-aware Set Transformer network; Section IV presents the experiment settings and results; limitations and a conclusion are given in Section V and Section VI, respectively.
	
	\section{Related Works}
	\subsection{Input features}
	Both time-domain and spectral-domain features have been widely used to analyze EEG signals. Time-domain features utilized by Craley et al.\cite{craley2019integrating} and Tsiouris et al.\cite{tsiouris2018long} include mean, variance, skewness, kurtosis, zero-crossing, etc. Although features extracted from the time-domain are intuitive for humans, they are more susceptible to noise and other disturbances. On the other hand, different cerebral rhythms are classified based on their frequency range in the medical field, and spectral features play vital roles in classification and prediction tasks based on EEG data. Time–frequency features can be extracted by Short Time Fourier Transform or Wavelet Transform in the form of spectrograms, and many studies, such as \cite{affes2022personalized,li2022eeg,truong2018convolutional,zhang2023distilling,cao2019epileptic,yuan2018novel,yuan2018multi}, chose spectrograms directly as their inputs. Differently, Zhang et al.\cite{zhang2015low} and Singh et al.\cite{singh2022two} selected the spectral power distribution over different frequency bands as their inputs. Additionally, Zeng et al.\cite{zeng2020hierarchy} and Li et al.\cite{li2020epileptic} combined time-domain and spectral-domain features together as inputs. There are also studies, such as Li et al.\cite{li2022eeg1}, that used raw EEG signals as inputs.
	
	Recently, research has been conducted to address the imbalance between pre-ictal and inter-ictal data. Li et al. \cite{li2024eeg} implemented a semi-supervised approach for seizure prediction, while Shu et al. \cite{shu2024data} employed a generative diffusion model for data augmentation in seizure prediction. Besides, Lopes et al. explored the effect of using a deep convolution neural network-based EEG artifact removal model \cite{lopes2023removing} and tried to addressing data limitations in seizure prediction through transfer learning \cite{lopes2024addressing}.
	\subsection{Deep learning algorithms}
	For works taking spectrograms as inputs, the convolutional neural network (CNN) is a typical solution for processing two-dimensional data and has been adopted by many studies\cite{affes2022personalized,li2022eeg,truong2018convolutional,zhang2023distilling,cao2019epileptic,yuan2018novel,yuan2018multi}. Some studies, such as Tsiouris et al.\cite{tsiouris2018long} and Singh et al.\cite{singh2022two}, utilized Long Short-Term Memory (LSTM) to further aggregate features.
	
	Transformer has outperformed RNN\cite{rumelhart1986learning} and its variants, such as LSTM and GRU\cite{cho2014learning}, in the natural language processing area in recent years, and it is one of the backbone algorithms of current popular large language models. Several solutions have been proposed to introduce Transformers into seizure prediction. Affes et al.\cite{affes2022personalized} and Hu et al.\cite{hu2023exploring} utilized transformer-based networks to complement a domain adversarial model and transfer learning. Li et al.\cite{li2022eeg} combined CNN and Transformer to extract features. Hussein et al.\cite{hussein2022multi} fed spectrograms directly to Transformer after position embedding, and Yan et al. \cite{yan2022seizure} transformed the spectrograms to three matrices before feeding them to Transformer. Furthermore, Koutsouvelis et al. \cite{koutsouvelis2024preictal} trained a competent subject-specific CNN-Transformer model to identify the optimal pre-ictal period for seizure prediction.
	
	In our study, we took a different approach by selecting several band power features as inputs and utilizing the Set Transformer to process these inputs temporally and channel-wisely. Set Transformer, compared to standard Transformer, decreases the computation complexity of self-attention from quadratic to linear in the number of elements. Therefore, in our study, the need for position embedding was eliminated, and computational overhead was reduced. 
	
	\begin{figure*}[bth]
		\centering
		\includegraphics[width=0.95\textwidth]{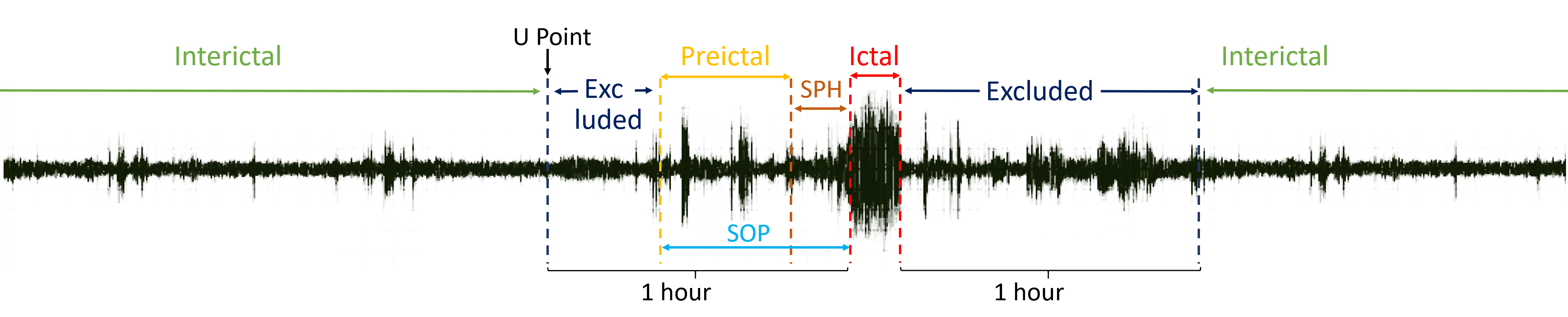}
		\caption{Illustrations of ictal/preictal/interictal periods, SPH, and data excluded. `U' point is at the beginning of an uncertain period.}
		\label{fig_def}
	\end{figure*}
	
	\subsection{Graph or channel selection}
	Some studies \cite{zeng2020hierarchy,zhao2021eeg,tsiouris2018long} employed graph neural networks (GCN) \cite{bruna2013spectral,defferrard2016convolutional} to aggregate signals from different channels. GCN values the connections between channels and builds Laplacian Matrices with Adjacency Matrices and Degree Matrices so that features can diffuse between channels. Conversely, some researchers hypothesized that the prediction results could be dominated by several channels.
	
	Reducing the number of EEG collecting sensors and covering a higher proportion of 24-hour EEG signals help introduce seizure prediction devices into the daily life of patients. Some studies \cite{birjandtalab2017automated,daoud2019efficient,karimi2020multi,ein2021automated,yuan2017multi} calculated different metrics for each channel and then selected a pre-defined number of channels whose metrics rank best. Zhang et al. \cite{zhang2015low} elaborated on patient-specific channel selection via iteration over all possible channel combinations and picked features through a branch and bound algorithm.
	
	Birjandtalab et al. \cite{birjandtalab2017automated} adopted a random forest ranking to pick channels. Affes et al.\cite{affes2022personalized} developed an unassisted methodology that identified and selected the most pertinent EEG channels for each patient without pre-defining any parameters, and this work only excluded data one-hour before or after seizures. Specifically, it designed a two-phase approach for epileptic seizure prediction: a neural network with an attention mechanism selected channels at the first phase, and another neural network processed data of the selected channels at the second phase. On the other hand, our two-stage Set Transformer network is capable of conducting seizure prediction and channel selection simultaneously.
	
	\section{Method}
	\subsection{Dataset and Configurations}
	The MIT Physionet EEG (CHB-MIT) dataset, acquired at Children’s Hospital Boston, encompasses EEG recordings from pediatric subjects with intractable seizures. The International 10–20 system \cite{acharya2016american}, which defines EEG electrode positions and nomenclature, is employed for these recordings. The dataset contains 23 subjects, with EEG recording durations spanning between 9 and 42 hours. We selected 18 channels shared by all the patients as input channels. Patient 23 was not included in this study because the subject only contains 0.27 hour interictal EEG data.
	
	Many prior studies \cite{affes2022personalized,li2022eeg,li2022eeg1,truong2018convolutional,zhang2023distilling} chose to leave out four-hour EEG signals before or after seizures to constitute the interictal data. However, adhering to this clinical division results in ambiguous prediction outputs during this period, likely leading to alarms being triggered four hours before a seizure. Considering the possibility of incorrect predictions, this provides limited assistance to patients in managing their schedules based on the predictive outcomes. To mitigate this issue, our study only excluded one-hour data before or after seizures to form the interictal data. Thus, even if an alarm is triggered at the beginning of the uncertain period, as the `U' point depicted in Figure \ref{fig_def}, a seizure will occur within one hour.
	
	The seizure prediction horizon (SPH) and seizure occurrence period (SOP) are defined by Maiwald et al. \cite{maiwald2004comparison}, as shown in Figure \ref{fig_def}. The SOP denotes the period in which a seizure is anticipated to occur, and the SPH should provide sufficient time for doctors to perform clinical interventions or for patients to implement safety precautions. For a correct prediction, a seizure onset must occur after the SPH and within the SOP. Conversely, a false alarm occurs when the predictive system gives a positive result but no seizure transpires during the SOP. In this study, the SPH was set to 3 minutes, and the SOP was set to 30 minutes. Moreover, if the interval between two consecutive seizures was less than one hour, the latter one was not treated as an independent seizure but was combined with the former one.
	
	\begin{figure*}[hbt]
		\centering
		\includegraphics[width=\textwidth]{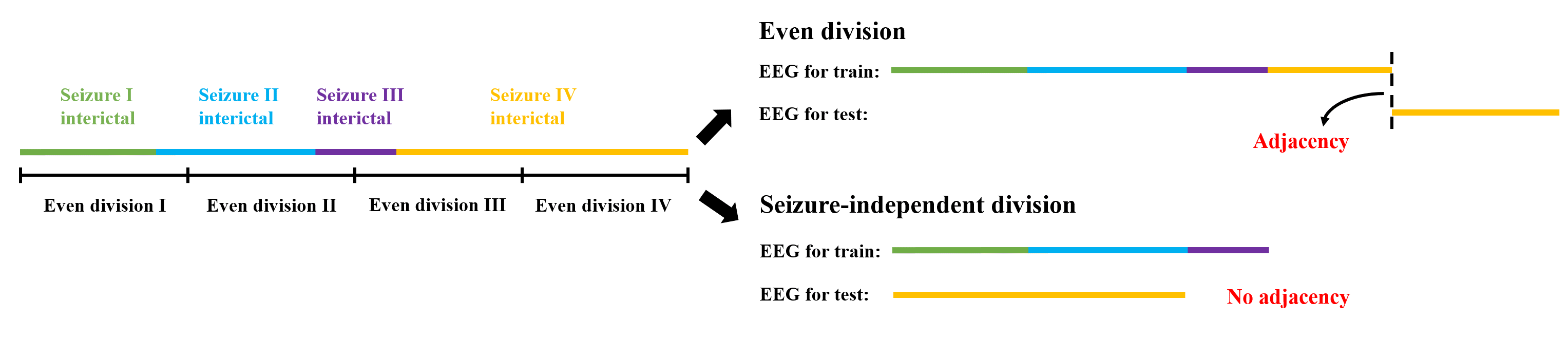}
		\caption{An illustration of the even division and the seizure-independent division.}
		\label{fig_division}
	\end{figure*}
	
	\begin{figure}[hbt]
		\centering
		\includegraphics[width=0.9\columnwidth]{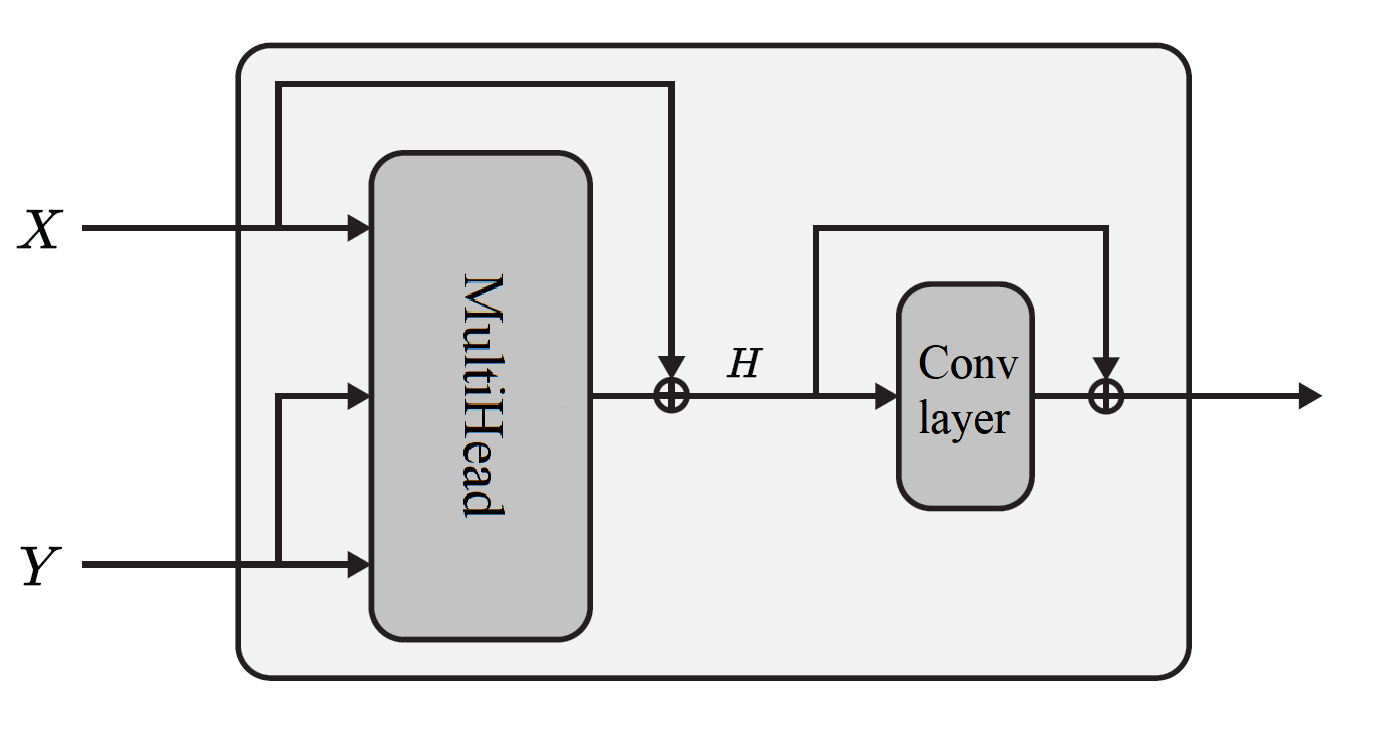}
		\caption{The architecture of MAB. $\oplus$ means adding operation, and \\$H=\operatorname{LayerNorm}(X+\operatorname{MultiHead_{MAB}}(X,Y))$}
		\label{fig_MAB}
	\end{figure}
	
	\subsection{Preprocessing}
	\label{sec_band} 
	An appropriate segment length should be set to make a trade off among temporal resolution, information integrity, and computational overhead. In this study, the EEG signals were analyzed in a 2-seconds-long segment with 50\% overlapping following Zeng et al.\cite{zeng2020hierarchy} and Cao et al.\cite{cao2019epileptic}.
	
	Each EEG signal segment was processed using a fast Fourier transform method \cite{welch1967use} to obtain its power spectra. EEG signals are usually accompanied by noise from sources like muscles, eye movements, sleep spindle waves, power lines, and the environment. To enhance the robustness of the entire system, non-physiological slow drifts, power line noise, and high-frequency environmental noise should be eliminated \cite{wu2020electroencephalographic}. Therefore, a band-stop filter was used to remove the spectral power with a frequency between 57 and 63 Hz, and a band-pass filter was used to discard spectral power with a frequency above 128 Hz.
	
	Several studies \cite{affes2022personalized,li2022eeg,truong2018convolutional,zhang2023distilling,cao2019epileptic,yuan2018novel,yuan2018multi} utilized spectrograms generated by wavelet transform as their inputs directly. Differently, following Zhang et al.\cite{zhang2015low}, we employed several band power features (absolute spectral power, relative spectral power, and spectral power ratio) as low-complexity inputs.
	
	The rhythmic activity in an EEG signal is typically described in terms of the standard frequency bands. In our study, the band was divided into eight sub-bands: (1) $\theta$(4–8 Hz), (2) $\alpha$(8–13 Hz), (3) $\beta$(13–30 Hz), (4) $\gamma$1(30–50 Hz), (5)  $\gamma$2(50–70 Hz), (6) $\gamma$3(70–90 Hz), (7) $\gamma$4(90–110 Hz), (8) $\gamma$5(110–128 Hz). The absolute spectral power of a signal in a frequency band is computed as the logarithm of the sum of the power spectral density (PSD) coefficients within that frequency band:
	\begin{equation}
		P_{absolute}(i)=log\sum_{w\in \text{band}\ i}PSD(w),\ i=1,2,...,8
	\end{equation}
	
	The relative spectral power measures the ratio of the power in the i-th band to the total power of the signal in a logarithmic scale, which is computed as:
	\begin{equation}
		P_{relative}(i)=log\frac{\sum_{w\in \text{band}\ i}PSD(w)}{\sum_{w\in \text{all band}}PSD(w)},\ i=1,2,...,8
	\end{equation}
	
	The spectral power ratios represent the spectral power ratio of every two bands, with the ratio of band i to band j represented as:
	\begin{equation}
		R_{i,j}=P_{absolute}(i)\ - \ P_{absolute}(j)
	\end{equation}
	Spectral power ratios have been effectively used as features in \cite{parhi2013seizure} for seizure prediction.
	
	In summary, 44 features, including eight absolute power features, eight relative power features, and 28 power ratio features, were extracted for each electrode channel.
	
	\subsection{EEG Sequence Division} 
	Comparing the performance of works with different EEG dividing schemes is not meaningful, as dividing schemes can seriously influence prediction results. LOOCV (leave one out cross validation) has been widely used to prevent overfitting. However, in previous LOOCV, the event left as testing data only included the preictal data of a seizure and did not include the interictal data before the seizure. For example, for a patient with n seizures, all the interictal records are concatenated together and then split evenly into n segments. Afterward, the n interictal segments and n preictal segments are merged respectively to form n pseudo-independent events.
	
	We refer to this division method as ``even division". It is widely implemented because many patients have no or only one seizure with its own independent interictal periods, such as Patient 3 and Patient 11 in Table \ref{tab_chb}, even if we only excluded one-hour data before or after a seizure. Most previous works excluded four-hour data before or after a seizure, which makes the phenomenon more serious. However, the even division may increase the risk of overfitting. It breaks an integrated interictal sequence before a specific seizure into pieces which can appear in both training and test data, as shown in Figure \ref{fig_division}. This results that machine learning methods may learn to memorize the features of adjacent EEG sequences rather than identifying the latent differences between preictal and interictal sequences. 
	
	To tackle this concern, our study employed a seizure-independent division, where interictal sequences can only be divided by seizures. This division avoids the adjacency of training and test data, as depicted in Figure \ref{fig_division}. Furthermore, the seizure-independent division aligns more closely with clinical practice, where the EEG sequence to be predicted is unknown, and therefore neither its preictal sequence nor its interictal sequence should be included during training.
	
	\begin{figure*}[hbt]
		\centering
		\includegraphics[width=0.9\textwidth]{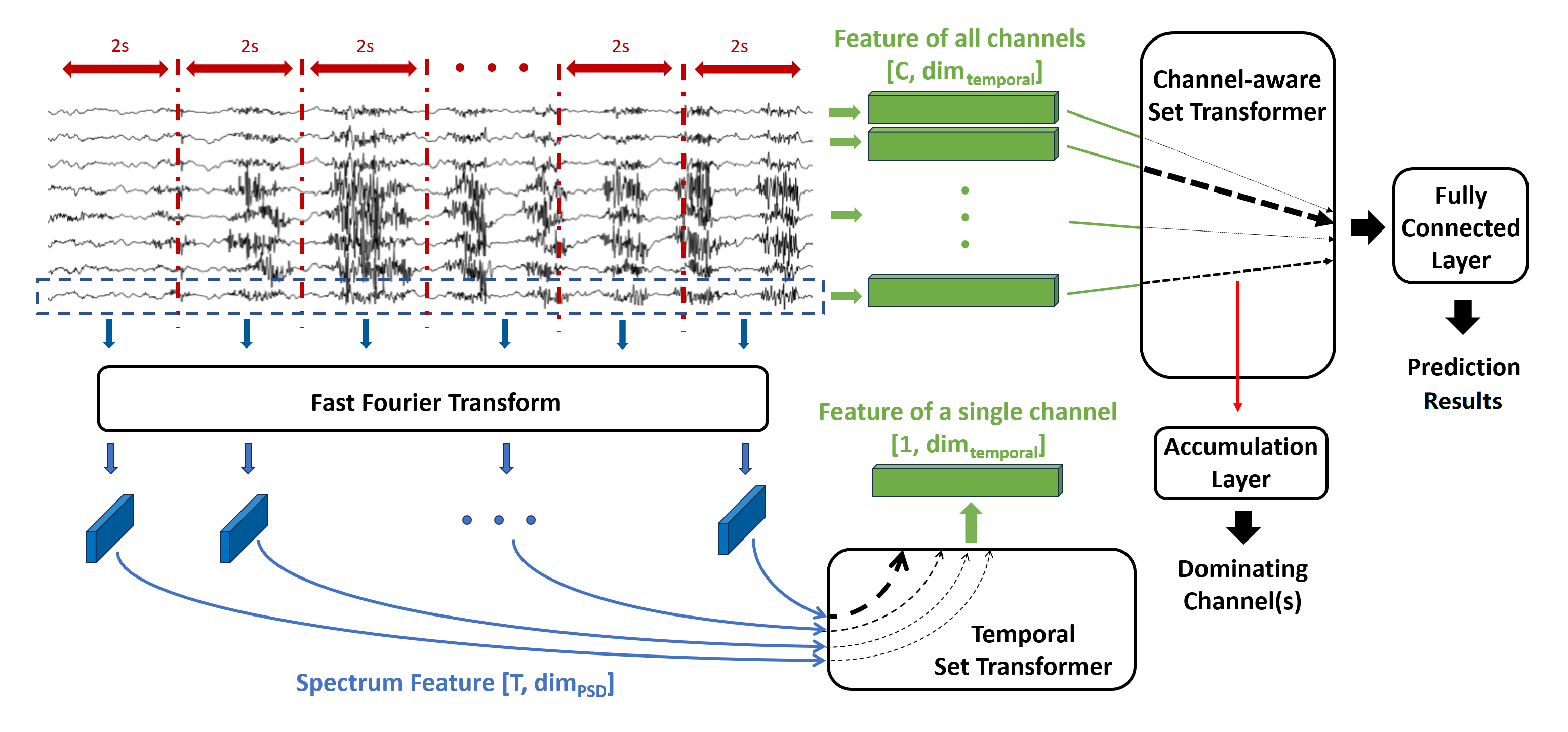}
		\caption{The overall architecture of the proposed two-step Set Transformer seizure predictor. T is the number of EEG signal segments, and C is the number of electrode channels.}
		\label{fig_transformer}
	\end{figure*}
	
	\subsection{Network Architecture}
	\subsubsection{Set Transformer}
	Raw EEG signals are essentially multichannel one-dimensional signals in the time domain. Processing these signals involves two steps: extracting features from each individual electrode channel and combining the outputs of all electrode channels. CNN can be utilized to accomplish these steps within a network \cite{truong2018convolutional}, where different electrode channels are represented as input channels of the CNN. Besides, some studies opted for LSTM to aggregate features from different electrode channels\cite{tsiouris2018long,singh2022two}. Given that the Transformer has surpassed traditional RNN algorithms in natural language processing, it has also been recently introduced to handle EEG features after embedding the features with position encoding \cite{hussein2022multi,affes2022personalized}.
	
	However, using Transformer with additional position encoding may render the network redundant, leading to substantial computation overhead. On the contrary, Set Transformer \cite{lee2019set} is permutation invariant and reduces the computation time of self-attention from quadratic to linear in the number of elements.
	
	The Scaled Dot-Product Attention in the Transformer can be computed as:
	\begin{equation}
		\operatorname{Attention}(Q, K, V)=\operatorname{softmax}\left(\frac{Q K^T}{\sqrt{d_k}}\right) V
	\end{equation}
	where $Q$, $K$ $\in \mathbb{R}^{1 \times d_k}$ represents queries and keys, respectively; $V$ $\in \mathbb{R}^{1 \times d_v}$ represents values.
	
	Multi-head attention version of attention is found more beneficial and applied widely in previous studies:
	\begin{equation}
		\operatorname{MultiHead}(Q, K, V) =\operatorname{Concat}\left({head}_1, \ldots, {head}_{\mathrm{h}}\right) W^O 
	\end{equation}
	where $head_i  =\operatorname{Attention}\left(Q W_i^Q, K W_i^K, V W_i^V\right)$, and $W^O$ is a matrix fusing the outputs of $h$ heads together. Specifically, $\operatorname{MultiHead}(\cdot, \cdot, \cdot)$ has learnable parameters $\left\{W_i^Q, W_i^K, W_i^V\right\}_{i=1}^h$, where $W_i^Q, W_i^K \in \mathbb{R}^{d_k \times d_k^M}$, $W_i^V \in \mathbb{R}^{d_v \times d_v^M}, W^O \in \mathbb{R}^{h d_v^M \times d}$; $d_k^M$ and $d_v^M$ are the dimensions of keys and values of a single head, respectively; and $d$ is the dimension of the output of  $\operatorname{Multihead}(\cdot, \cdot, \cdot)$. In this study, we set $d_k^M=d_k / h, d_v^M=d_v / h$ following Lee et al. \cite{lee2019set}.
	
	In Set Transformer, a Multihead Attention Block (MAB) is designed to process permutation invariant inputs. In the attention of MAB, the inputs of keys are the same as the inputs of values:
	\begin{equation}
		\begin{aligned}
			&\operatorname{MultiHead_{MAB}}(X,Y)\\
			=&\operatorname{MultiHead}(\operatorname{Conv_Q}(X), \operatorname{Conv_K}(Y), \operatorname{Conv_V}(Y))
		\end{aligned}
	\end{equation}
	
	After adding feed-forward layers, the MAB can be computed as:
	\begin{equation}
		\operatorname{MAB}(X, Y)=\operatorname{LayerNorm}\left(H+\operatorname{Relu}(\operatorname{Conv_H}(H))\right)\\
	\end{equation}
	where $H=\operatorname{LayerNorm}(X+\operatorname{MultiHead_{MAB}}(X,Y))$. The architecture of MAB is illustrated in Figure \ref{fig_MAB}.
	
	The Set Transformer's inherent permutation invariance property aligns with our observation that certain segments of the EEG signal and certain channels hold greater predictive importance than others, and the specific order in which they are arranged is less significant. It allows us to input features from different time points simultaneously in the temporal Set Transformer at the first stage, without the need for individual position encoding for each input. Similarly, in the second stage, features from different EEG electrode channels can be processed without the need for order encoding. 
	
	The overall architecture of the two-stage channel-aware Set Transformer is illustrated in Figure \ref{fig_transformer}.
	\subsubsection{Temporal Set Transformer}
	Inputs of the two-stage Set Transformer were EEG sequences of 38 seconds. The PSD features were extracted every 2 seconds so that features of 19 EEG signal segments were fed into the network at a time for each channel. Since EEG data typically fluctuates rapidly, predictions based on a relatively longer period of time tend to be more robust. Furthermore, we observed that certain segments of the EEG signal have greater importance for prediction than others, and the specific order in which these segments are inputted is less relevant. Thus, we utilized a Set Transformer in the first stage to assign attention to the temporal features and merge them by introducing a temporal kernel:
	\begin{equation}
		\begin{aligned}
			&{Feature}_{\text{channel}_i}
			=\operatorname{MAB}(Kernel_\text{temp}, PSD_{\text{channel}_i})
		\end{aligned}
	\end{equation}
	where $Kernel_\text{temp} \in \mathbb{R}^{1 \times dim_\text{temporal}}$ is a randomly initialized tensor which is trainable, and $dim_\text{temporal}$ is the dimension of the temporal kernel tensor. $PSD_{\text{channel}_i} \in \mathbb{R}^{T \times dim_\text{PSD}}$ is the band power features of channel $i$. $T$ is the number of EEG signal segments fed into the network at one time and was 19 in this study, and $dim_\text{PSD}$ is the dimension of the PSD features and was 44 as described in Section \ref{sec_band}. ${Feature}_{\text{channel}_i}\in \mathbb{R}^{dim_\text{temporal}}$ is the output feature of the i th channel.
	
	\begin{figure*}[hbt]
		\centering
		\includegraphics[width=0.8\textwidth]{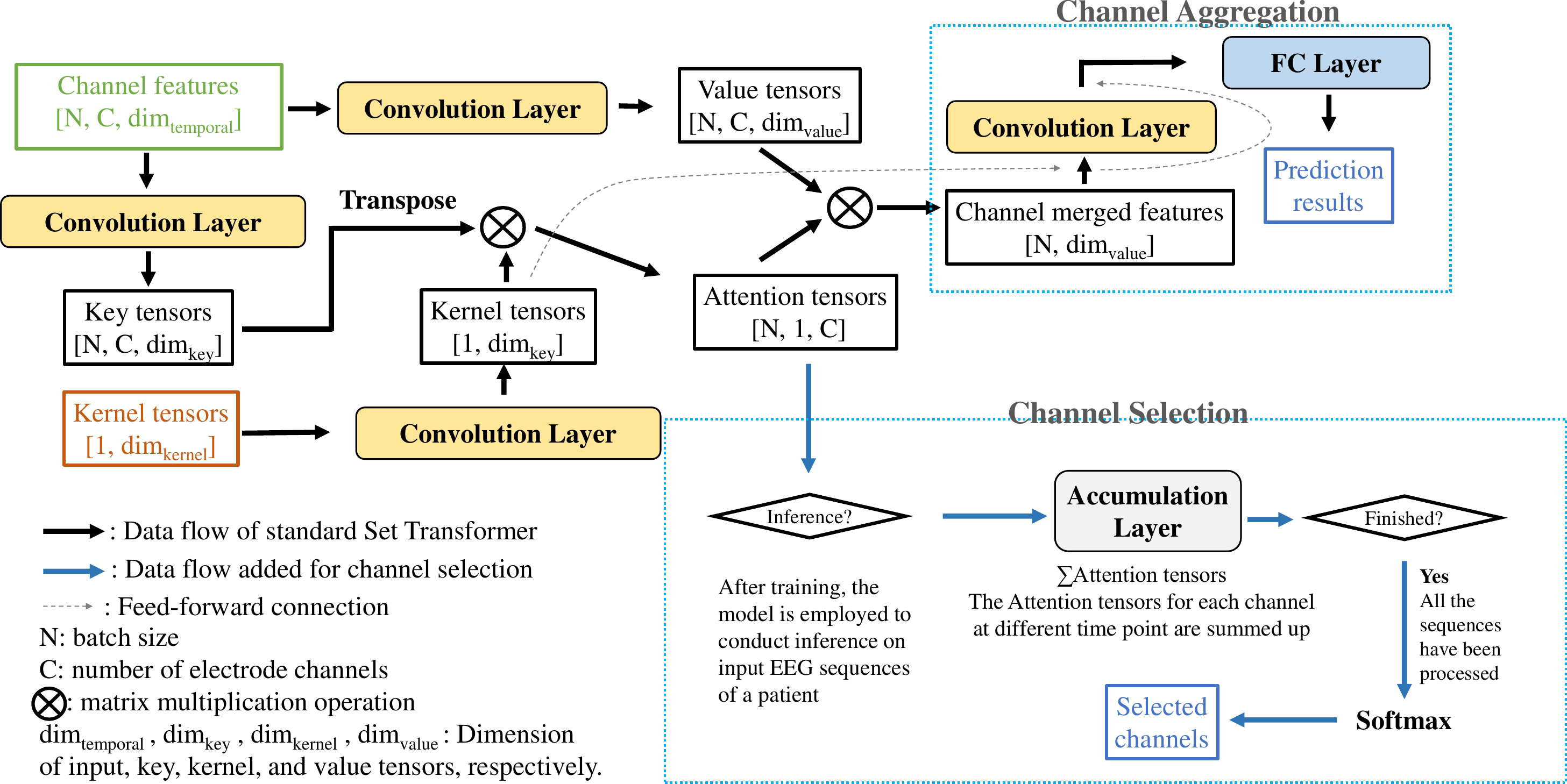}
		\caption{The data flow of the channel-aware Set Transformer.}
		\label{fig_aware}
	\end{figure*}
	
	\subsubsection{Channel-Aware Set Transformer}
	In the second stage, the temporally merged features from all channels were further processed by a channel-aware Set Transformer:
	\begin{equation}
		\begin{aligned}
			&{Feature}_\text{output} \\
			=&\operatorname{MAB}_\text{CHAW}(Kernel_\text{channel}, Feature_\text{channel})
		\end{aligned}
	\end{equation}
	where $\operatorname{MAB}_\text{CHAW}$ is the channel-aware Set Transformer; $Kernel_\text{channel} \in \mathbb{R}^{1 \times dim_\text{output}}$ is a randomly initialized tensor which is trainable; $Feature_\text{channel} \in \mathbb{R}^{C \times dim_\text{temporal}}$ is the output feature of the first stage, and $C$ represents the number of electrode channels and was 18 in this study; ${Feature}_\text{output}\in \mathbb{R}^{dim_\text{output}}$ is the output of the second stage and will be inputted to a fully connected layer to get the final predictive results. 
	
	The attentions assigned to different channels fluctuate fiercely among different batches during training and inference. As a result, an accumulation layer was supplemented to a Set Transformer to form the channel-aware Set Transformer, $\operatorname{MAB}_\text{CHAW}$. This layer accumulated the attention values batch by batch during inference, and the accumulations were fed to a Softmax function to conduct channel selection after finishing the inference for all inputs:
	\begin{equation}
		{Attention}_{acc}=\operatorname{Softmax}\left(\frac{\sum^{N_{batch}}\sum^{N}{Attention}_{seq}}{N_{batch}\times N}\right)
	\end{equation}
	where ${Attention}_{seq} \in \mathbb{R}^{1\times C}$ represents the attention distribution of a single input; $N$ is the batch size, and the input EEG sequence is divided into $N_{batch}$ batches; ${Attention}_{acc}\in \mathbb{R}^{1\times C}$ represents the attention among channels after the accumulation.
	
	The data flow of the channel-aware Set Transformer is illustrated in Figure \ref{fig_aware}.
	
	Patient-specific channel selection can be realized if the prediction results of a patient are dominated by a few channels. On the other hand, the channel selection is to fail if no dominant channels appear. If the attentions of channels succeed to converge, channels with the most significant attentions will be chosen as the dominant channels. Reducing the number of electrode channels can lead to smaller, less energy-consuming EEG-collecting devices, increasing their acceptance among epilepsy patients.
	
	After the channel selection, only EEG signals from those dominant channels were fed to our two-stage Set Transformer again as inputs. Retraining was needed after channel selection because of the change of input data. Given the architecture of the Set Transformer, no modification was needed for the network to process inputs with fewer channels.
	
	\begin{figure*}[hbt]
		\centering
		\includegraphics[width=0.9\textwidth]{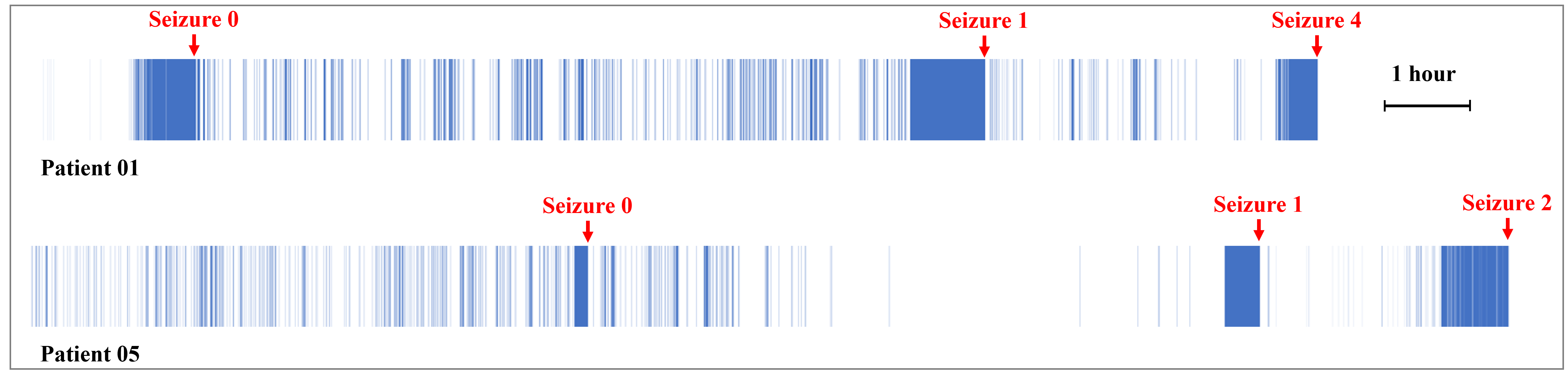}
		\caption{Visualization of outputs of the predictor during interictal and preictal periods for Patient 01 and Patient 05. Most false-positive predictions during interictal periods can be filtered because they do not last long.}
		\label{fig_output}
	\end{figure*}
	
	\section{Results}
	\subsection{Settings and Metrics}
	Experiments were conducted on the EEG recordings of 22 patients from the CHB-MIT scalp EEG dataset. Adjacent seizures with an interval of less than one hour were merged as one seizure. Eighteen channels shared by the patients were selected as input channels: FP1-F7, F7-T7, T7-P7, P7-O1, P3-O1, C3-P3, F3-C3, FP1-F3, FZ-CZ, CZ-PZ, P4-O2, C4-P4, F4-C4, FP2-F4, FP2-F8, F8-T8, T8-P8, and P8-O2, which were indexed from 0 to 17, respectively.
	
	In our study, one-hour data before or after seizures were excluded from the analysis. The SPH and SOP were set to 3 minutes and 30 minutes, respectively. A refractory period of 30 minutes was established to prevent continuous triggering of alarms within a short timeframe. Some studies \cite{truong2018convolutional,zhang2023distilling} employed a k-of-n approach to mitigate the impact of isolated false-positive predictions during interictal periods. Specifically, an alarm would be triggered if at least k positive predictions occurred within n consecutive windows. In our study, an alarm would be activated if positive predictions continued for a certain duration without any interjected negative predictions. A threshold of approximately 4 minutes was suitable for most patients in our study, while a threshold of 1 or 2 minutes worked better for several patients.
	
	Existing seizure prediction tasks are classified into two categories: segment-based and event-based. Our study employed event-based prediction. Sensitivity (Sen) and false predicting rate (FPR) were chosen as the primary metrics for evaluating model performance. Sensitivity refers to the percentage of seizures correctly predicted, and false predicting rate represents the average number of false alarms raised in one hour.
	\begin{equation}
		\text{Sen}  =\text{TP}/(\text{FN} + \text{TP}) \times 100 \% 
	\end{equation}
	
	\begin{equation}
		\text{FPR}=\text{FP}/{\operatorname{Length}_\text{interictal}}
	\end{equation}
	where TP is the number of correctly predicted seizures, FN is the number of seizures models fail to predict, FP is the number of wrong alarms raised during interictal, and $\operatorname{Length}_\text{interictal}$ means the total length of interictal EEG signals.
	
	Experiments were conducted on both the even division and the seizure-independent division.
	\begin{table}[tbh]
		\centering
		\caption{The predicting results when the even division is conducted}
		\label{tab_even_division}
		\begin{tabular}{cccccc}
			\hline
			\multicolumn{1}{c|}{\multirow{2}{*}{ID}} & \multicolumn{2}{c|}{All channel}     & \multicolumn{3}{c}{After channel   selection}                                                       \\ \cline{2-6} 
			\multicolumn{1}{c|}{}                    & FPR/h & \multicolumn{1}{c|}{Sensitivity}  & \begin{tabular}[c]{@{}c@{}}Selected\\      channel(s)\end{tabular}                 & FPR/h & Sensitivity  \\ \hline
			\multicolumn{1}{c|}{01}                  & 0.14  & \multicolumn{1}{c|}{100.0\%} & 12,14                                                                             & 0.07  & 100.0\% \\
			\multicolumn{1}{c|}{02}                  & 0.13  & \multicolumn{1}{c|}{66.7\%}  & 9,12                                                                              & 0.07  & 100.0\% \\
			\multicolumn{1}{c|}{03}                  & 0.25  & \multicolumn{1}{c|}{80.0\%}  & 6,9                                                                               & 0.00  & 60.0\%  \\
			\multicolumn{1}{c|}{04}                  & 0.09  & \multicolumn{1}{c|}{66.7\%}  & 3,5,9                                                                               & 0.05  & 33.3\%  \\
			\multicolumn{1}{c|}{05}                  & 0.00  & \multicolumn{1}{c|}{100.0\%} & 8,9,10                                                                            & 0.00  & 100.0\% \\
			\multicolumn{1}{c|}{06}                  & 0.10  & \multicolumn{1}{c|}{70.0\%}  & 3,9,13,17                                                                         & 0.10  & 70.0\%  \\
			\multicolumn{1}{c|}{07}                  & 0.03  & \multicolumn{1}{c|}{33.3\%}  & 12,13,16,17                                                                         & 0.06  & 33.3\%  \\
			\multicolumn{1}{c|}{08}                  & 0.00  & \multicolumn{1}{c|}{80.0\%}  & 3,14                                                                              & 0.00  & 80.0\%  \\
			\multicolumn{1}{c|}{09}                  & 0.18  & \multicolumn{1}{c|}{0.0\%}  & 3,11,17                                                                              & 0.20  & 66.7\%  \\
			\multicolumn{1}{c|}{10}                  & 0.06  & \multicolumn{1}{c|}{83.3\%}  & 4,9                                                                               & 0.03  & 83.3\%  \\
			\multicolumn{1}{c|}{11}                  & 0.06  & \multicolumn{1}{c|}{100.0\%} & 0,4,10                                                                            & 0.03  & 100.0\% \\
			\multicolumn{1}{c|}{12}                  & 0.00  & \multicolumn{1}{c|}{100.0\%} & 8,9,16                                                                            & 0.25  & 100.0\% \\
			\multicolumn{1}{c|}{13}                  & 0.00  & \multicolumn{1}{c|}{75.0\%}  & 6                                                                                 & 0.00  & 75.0\%  \\
			\multicolumn{1}{c|}{14}                  & 0.19  & \multicolumn{1}{c|}{60.0\%}  & 4,5,8,9                                                                           & 0.74  & 100.0\% \\
			\multicolumn{1}{c|}{15}                  & 0.00  & \multicolumn{1}{c|}{100.0\%}  & 2,3,7                                                                           & 0.11  & 100.0\% \\
			\multicolumn{1}{c|}{16}                  & 0.14  & \multicolumn{1}{c|}{100.0\%} & \multicolumn{3}{c}{Failed}                                                                          \\
			\multicolumn{1}{c|}{17}                  & 0.16  & \multicolumn{1}{c|}{100.0\%} & 3,4,8,17                                                                          & 0.32  & 100.0\%\\
			\multicolumn{1}{c|}{18}                  & 0.00  & \multicolumn{1}{c|}{50.0\%}  & 1,6                                                                               & 0.00  & 50.0\%  \\
			\multicolumn{1}{c|}{19}                  & 0.07  & \multicolumn{1}{c|}{100.0\%} & 2,6,8,16,17                                                                       & 0.00  & 100.0\% \\
			\multicolumn{1}{c|}{20}                  & 0.15  & \multicolumn{1}{c|}{75.0\%}  & 8,14                                                                              & 0.05  & 75.0\%  \\
			\multicolumn{1}{c|}{21}                  & 0.06  & \multicolumn{1}{c|}{75.0\%}  & 8                                                                                 & 0.12  & 75.0\%  \\
			\multicolumn{1}{c|}{22}                  & 0.30  & \multicolumn{1}{c|}{66.7\%}  & \multicolumn{3}{c}{Failed}                                                                          \\ \hline
			\multicolumn{1}{c|}{Mean}                & 0.09  & \multicolumn{1}{c|}{76.4\%}  & \multirow{2}{*}{\begin{tabular}[c]{@{}c@{}}2.8 channels\\ in average\end{tabular}} & 0.11  &  80.1\% \\
			\multicolumn{1}{c|}{All}                 & 0.09  & \multicolumn{1}{c|}{76.1\%}  &                                                                                   & 0.07  &  79.3\% \\ \hline
			\multicolumn{6}{l}{\multirow{2}{*}{\begin{tabular}[c]{@{}l@{}}Mean: The metric of every patient is calculated individually. After\\that, an average of all patients is computed.\\ All: Seizures from different patients are grouped together to calculate\\the metrics.\end{tabular}}}                                    \\
			\multicolumn{6}{l}{}                                                                                                                                                                 
		\end{tabular}
	\end{table}
	
	\begin{table}[bth]
		\centering
		\caption{The length of interictal and preictal before each seizure in CHB-MIT dataset}
		\label{tab_chb}
		\begin{tabular}{cclllllll}
			\hline
			ID                  & \multicolumn{1}{l}{Period} & S0   & S1   & S2  & S3   & S4  & S5   & … \\ \hline
			\multirow{2}{*}{\textbf{1}}  & Interictal                 & 1.8  & 9.1  & 0.2 & 0.4  & 3.4 &      &   \\
			& Preictal                   & 0.5  & 0.5  & 0.5 & 0.4  & 0.5 &      &   \\ \hline
			\multirow{2}{*}{2}  & Interictal                 & 14.0 & 0.0  & 1.1 &      &     &      &   \\
			& Preictal                   & 0.5  & 0.5  & 0.5 &      &     &      &   \\ \hline
			\multirow{2}{*}{3}  & Interictal                 & 0.0  & 0.0  & 0.0 & 28.4 & 0.0 & 0.0  & … \\
			& Preictal                   & 0.1  & 0.5  & 0.5 & 0.5  & 0.5 & 0.5  & …  \\ \hline
			\multirow{2}{*}{\textbf{5}}  & Interictal                 & 4.1  & 5.2  & 1.3 & 0.0  & 2.9 &      &   \\
			& Preictal                   & 0.5  & 0.5  & 0.5 & 0.5  & 0.5 &      &   \\ \hline
			\multirow{2}{*}{\textbf{6}}  & Interictal                 & 0.0  & 0.0  & 0.0 & 6.3  & 0.0 & 19.4 &…   \\
			& Preictal                   & 0.5  & 0.5  & 0.5 & 0.5  & 0.5 & 0.5  &…   \\ \hline
			\multirow{2}{*}{8}  & Interictal                 & 0.0  & 1.0  & 0.8 & 0.0  & 5.9 &      &   \\
			& Preictal                   & 0.5  & 0.5  & 0.5 & 0.5  & 0.5 &      &   \\ \hline
			\multirow{2}{*}{11} & Interictal                 & 32.0 & 0.0  & 0.0 &      &     &      &   \\
			& Preictal                   & 0.1  & 0.5  & 0.4 &      &     &      &   \\ \hline
			\multirow{2}{*}{\textbf{13}} & Interictal                 & 15.6 & 0.0  & 5.8 & 1.0  & 0.3 & 0.0  &…   \\
			& Preictal                   & 0.5  & 0.3  & 0.5 & 0.1  & 0.5 & 0.5  &…   \\ \hline
			\multirow{2}{*}{\textbf{14}} & Interictal                 & 1.6  & 0.0  & 0.5 & 3.4  & 5.8 &      &   \\
			& Preictal                   & 0.5  & 0.5  & 0.5 & 0.5  & 0.5 &      &   \\ \hline
			\multirow{2}{*}{16} & Interictal                 & 8.6  & 1.2  & 0.0 &      &     &      &   \\
			& Preictal                   & 0.5  & 0.5  & 0.5 &      &     &      &   \\ \hline
			\multirow{2}{*}{\textbf{18}} & Interictal                 & 28.0 & 0.0  & 0.0 & 2.0  & 0.0 &      &   \\
			& Preictal                   & 0.5  & 0.5  & 0.5 & 0.5  & 0.1 &      &   \\ \hline
			\multirow{2}{*}{19} & Interictal                 & 27.0 & 0.0  & 0.0 &      &     &      &   \\
			& Preictal                   & 0.1  & 0.5  & 0.5 &      &     &      &   \\ \hline
			\multirow{2}{*}{20} & Interictal                 & 8.0  & 0.0  & 0.0 & 12.4 &     &      &   \\
			& Preictal                   & 0.5  & 0.5  & 0.5 & 0.4  &     &      &   \\ \hline
			\multirow{2}{*}{21} & Interictal                 & 17.4 & 0.0  & 0.0 & 0.0  &     &      &   \\
			& Preictal                   & 0.5  & 0.5  & 0.5 & 0.5  &     &      &   \\ \hline
			\multirow{2}{*}{\textbf{22}} & Interictal                 & 15.9 & 2.9  & 4.1 &      &     &      &   \\
			& Preictal                   & 0.5  & 0.5  & 0.4 &      &     &      &   \\ \bottomrule
			\multicolumn{9}{l}{{\begin{tabular}[c]{@{}l@{}}ID: The index of patients. If an ID is in bold, the patient was included\\in the experiment of seizure-independent division.\\Si: The index of seizures.\\The unit of the length is hour.\end{tabular}}}  
		\end{tabular}
	\end{table}
	
	\begin{table*}[htb]
		\centering
		\caption{Comparison with other studies}
		\label{overall_acc}
		\begin{tabular}{ccccccc}
			\toprule
			Works                                                                                  & \begin{tabular}[c]{@{}l@{}}Channel selection\\ method\end{tabular}                                  & \begin{tabular}[c]{@{}l@{}}Number of\\ patients\end{tabular} & \begin{tabular}[c]{@{}l@{}}Excluded length \\ before/after \\ seizures\end{tabular} & \begin{tabular}[c]{@{}l@{}}Channel\\ number\end{tabular} & \begin{tabular}[c]{@{}l@{}}Sen\\ /\%\end{tabular} & FPR    \\
			\hline
			Truong et al.\cite{truong2018convolutional}                          & \multirow{2}{*}{\begin{tabular}[c]{@{}l@{}}No channel selection\\ No channel selection\end{tabular}}      & 13                                                           & 4 hour                                                                              & All                                                       & 81.2                                                 & 0.16/h \\ \hline
			Zhang et al.$\tiny{^*}$\cite{zhang2023distilling}                              &                                                                                                    & 13                                                           & 4 hour                                                                              & 18                                                       & 79.5                                                 & 0.26/h \\ \hline
			\multirow{2}{*}{\begin{tabular}[c]{@{}c@{}}Shu et al.\cite{shu2024data} without DA\\Shu et al. with DA\end{tabular}}   & \multirow{2}{*}{No channel selection}         & 13       & \multirow{2}{*}{-}    & All      & 90.7       & 0.103/h \\
			&       & 13         &          & All           & 95.4     & 0.062/h \\ \hline
			Our work                       & \multirow{1}{*}{\begin{tabular}[c]{@{}l@{}}Two-stage Set Transformer\end{tabular}}      & 13                                                           & 1 hour                                                                              & 2.8$\tiny{^{\#}}$                                                       & 95.3                                                 & 0.129/h \\ \bottomrule
			\multirow{2}{*}{Birjandtalab et al.\cite{birjandtalab2017automated}} & \multirow{2}{*}{Random Forest ranking}                                                             & \multirow{2}{*}{23}                                          & \multirow{2}{*}{-}                                                                  & All                                                       & 89.8                                                 & 2.20/h \\ 
			&                                                                                                    &                                                              &                                                                                     & 3                                                        & 80.9                                                 & 2.50/h \\ \hline
			\multirow{2}{*}{Abir Affes et al.\cite{affes2022personalized}}       & \multirow{2}{*}{\begin{tabular}[c]{@{}l@{}}Neural network with\\ attention mechanism\end{tabular}} & \multirow{2}{*}{23}    & \multirow{2}{*}{1 hour}        & 23         & 68.6     & 0.30/h \\
			&          &        &         & 2$^\#$             & 78.9      & 0.35/h \\ \hline
			\multirow{2}{*}{Our work}    & \multirow{2}{*}{Two-stage Set Transformer}         & 22       & \multirow{2}{*}{1 hour}    & 18      & 76.4       & 0.088/h \\
			&       & 20         &          & 2.8$\tiny{^{\#}}$           & 80.1     & 0.107/h \\ \bottomrule
			\multicolumn{7}{l}{\multirow{2}{*}{\begin{tabular}[c]{@{}l@{}}$\tiny{^*}$: It is a domain adaptation study.\\$\tiny{^\#}$: The average number of the selected channels.\\DA: data augmentation. \end{tabular}}}                                                                                                                                                                                                                                                                                                        
		\end{tabular}
	\end{table*}
	\subsection{Prediction Results on Even Division}
	To assess the prediction and channel selection capabilities of the two-stage channel-aware Set Transformer, experiments were carried out on the even division for 22 patients. The results are presented in Table \ref{tab_even_division}.
	
	Before channel selection, the mean sensitivity and mean false predicting rate of the patients were 76.4\% and 0.09 per hour, respectively. The overall sensitivity was 76.1\% with a false predicting rate of 0.09 per hour. Perfect sensitivity rates of 100\% were achieved for 8 patients, and the lowest sensitivity was zero for Patient 09.
	
	After introducing channel selection, dominant channels emerged in 20 out of the 22 patients. In other words, the channel attentions of Patient 16 and Patient 22 failed to converge. For the 20 patients, the mean sensitivity was 80.1\% with an FPR of 0.11 per hour. The overall sensitivity was 79.3\% with a false predicting rate of 0.07 per hour. The seizures of 9 patients were all accurately predicted. Patient 04 and Patient 07 had the lowest sensitivity of 33.3\%. The average number of selected channels was reduced to 2.8 from 18, which would allow EEG collecting devices to be smaller and consume less power, thus increasing their accessibility among epilepsy patients. A single dominant channel was sufficient for two (10.0\%) patients, seven (35.0\%) patients required two channels, three channels were needed for six (30.0\%) patients, four (20.0\%) patients demanded four channels, and Patient 19 (5.0\%) required the maximum number of five channels.
	
	\subsection{Prediction Results on the Seizure-Independent Division} 
	Following Zhang et al. \cite{zhang2015low}, only patients with a daily seizure frequency of more than 2 and less than 10 were included in the seizure-independent division. As shown in Table \ref{tab_chb}, several patients have only one or two independent interictal periods. Given that the seizure-independent division is more rigorous and the length of interictal periods for different seizures varies drastically, only seven patients with three or more seizure-independent interictal periods were analyzed. Before channel selection, the mean sensitivity and mean false predicting rate of the patients were 72.6\% and 0.08 per hour, respectively.
	After channel selection, the mean sensitivities remained the same, while the mean FPR increased to 0.10 per hour. 
	The predictive outputs for Patient 01 and Patient 05 are visualized in Figure \ref{fig_output} by concatenating the outputs during interictal and preictal periods. 
	
	The mean sensitivity for the same patients in the even division before channel selection was 79.3\% 
	with an FPR of 0.11 per hour. The sensitivity was 6.7\% higher than the results in the seizure-independent division. As channel selection failed for Patient 22 in the even division, the metrics after channel selection were computed among the remaining six patients, and the mean sensitivity in the even division was 88.1\% with an FPR of 0.15 per hour. 
	The discrepancies between the two division methods underscore our concerns about the traditional even division.
	
	\subsection{Comparison with Other Studies}
	Table \ref{overall_acc} provides a comparison between our channel selection method for epileptic seizure prediction and some previous studies that used the same dataset. The discussion is limited to results from the even division, as there are few studies, to our knowledge, that have performed seizure-independent divisions other than ours.
	
	
	Among studies without channel selection, Truong et al. \cite{truong2018convolutional} and Zhang et al. \cite{zhang2023distilling} excluded four-hour EEG data around seizures to create interictal datasets, likely making their data more distinguishable than ours, which excluded only one hour. Despite this, our sensitivity post-channel selection remains better with lower FPR. Besides, our results outperform Shu et al. \cite{shu2024data} before data augmentation and remain competitive afterward, despite using significantly fewer channels, and our performance could be further improved after introducing data augmentation methods.
	
	Birjandtalab et al. \cite{birjandtalab2017automated} and Affes et al. \cite{affes2022personalized} also introduced channel selection methods. Birjandtalab et al. achieved the highest sensitivity with the maximum false predicting rate, but its performance declined significantly after channel selection. Affes et al. utilized fewer channels than ours after selection and can be generalized to all patients. Nevertheless, our sensitivity and FPR outperformed those of Affes et al.
	
		\begin{table*}[htb]
		\centering
		\caption{Results of the ablation study}
		\label{tab_ablation}
		\begin{tabular}{ccccccccc}
			\hline
			\multirow{2}{*}{\begin{tabular}[c]{@{}c@{}}Solu-\\tion\end{tabular}} & \multirow{2}{*}{\begin{tabular}[c]{@{}c@{}}Division\\ method\end{tabular}}                 & \multirow{2}{*}{\begin{tabular}[c]{@{}c@{}}Input\\ features\end{tabular}}      & \multicolumn{2}{c}{Network type}                                                                                                    & \multicolumn{2}{c}{\begin{tabular}[c]{@{}c@{}}Metrics before\\ channel selection\end{tabular}} & \multicolumn{2}{c}{\begin{tabular}[c]{@{}c@{}}Metrics after\\ channel selection\end{tabular}} \\ \cline{4-9} 
			&                                                                                            &                                                                                & Temporal stage                                            & Channel selection stage                                                 & Mean Sen/\%                                     & Mean FPR/h                                     & Mean Sen/\%                                     & Mean FPR/h                                    \\ \hline
			1                         & \multirow{3}{*}{\begin{tabular}[c]{@{}c@{}}\\Even\\ division\end{tabular}}                   & \multirow{2}{*}{Spectrograms}                                                  & \begin{tabular}[c]{@{}c@{}}Depth-wise\\ CNN*\end{tabular} & \begin{tabular}[c]{@{}c@{}}Channel-aware\\ Set Transformer\end{tabular} & 74.8                                            & 0.11                                           & 78.5                                            & 0.14                                          \\ \cline{4-9} 
			2\cite{affes2022personalized}&                                                                                            &                                                                                & \begin{tabular}[c]{@{}c@{}}Depth-wise\\ CNN\end{tabular}  & CAtt-MLP + GRU                                                          & 68.6                                            & 0.30                                           & 78.9                                            & 0.35                                          \\ \cline{3-9} 
			3(Ours)                      &                                                                                            & \begin{tabular}[c]{@{}c@{}}Band power\\ features\end{tabular}                  & Set Transformer                                           & \begin{tabular}[c]{@{}c@{}}Channel-aware\\ Set Transformer\end{tabular} & 76.4                                            & 0.09                                           & 80.1                                            & 0.11                                          \\ \hline
			4                         & \multirow{2}{*}{\begin{tabular}[c]{@{}c@{}}Seizure-\\ independent\\ division\end{tabular}} & \multirow{2}{*}{\begin{tabular}[c]{@{}c@{}}Band power\\ features\end{tabular}} & LSTM                                                      & \begin{tabular}[c]{@{}c@{}}Channel-aware\\ Set Transformer\end{tabular} & 70.2                                            & 0.11                                           & 70.2                                            & 0.10                                          \\ \cline{4-9} 
			5(Ours)                      &                                                                                            &                                                                                & Set Transformer                                           & \begin{tabular}[c]{@{}c@{}}Channel-aware\\ Set Transformer\end{tabular} & 72.6                                            & 0.08                                           & 72.6                                            & 0.10                                          \\ \hline
			\multicolumn{9}{l}{\begin{tabular}[c]{@{}l@{}}*: Because the code of Affes et al.\cite{affes2022personalized} has not been released, the Depth-wise CNN was implemented by ourselves. Thus, it is to be different from that\\of Affes et al.\end{tabular}}
		\end{tabular}
	\end{table*}
	\subsection{Ablation study}
	Ablation studies were conducted to assess the contribution of different modules, and the results are presented in Table \ref{tab_ablation}. It would be unfair to compare channel selection methods that differ in their preceding stages or inputs. Therefore, we implemented Solution 1 of which the inputs and temporal stage resembled those of Affes et al. \cite{affes2022personalized}. Thereafter, the primary difference between Solution 1 and Affes et al. lay in the channel selection stage, enabling a relatively fair comparison. Subsequently, Solution 1 was evaluated under the even division. The mean sensitivity of Solution 1 was close to that of Affes et al., but its FPR was significantly lower, suggesting the effectiveness of our channel-aware Set Transformer module. Additionally, when compared to Solution 1, the original solution (Solution 3) achieved higher mean sensitivity with lower FPR, supporting the advantages of our selection regarding input features and temporal stages.
	
	Furthermore, in Solution 4, we replaced the original temporal stage with an LSTM network to evaluate the contribution of the Set Transformer in the first stage. The solution was evaluated under the seizure-independent division, and it was outperformed by the original solution (Solution 5). Nevertheless, the performance metrics of the two solutions were close with each other, inferring the superiority of the Set Transformer in the first stage may not be significant.
	
	\subsection{Model Complexity and Computational Time}
	Our model only contains 37.4K parameters, and 8.23M floating point operations are needed to conduct an inference. 590M GPU memory was occupied when the batch size was set to 16. The server we used for this study was equipped with a "GeForce GTX 1080 Ti" GPU and an "Intel(R) Xeon(R) CPU E5-2640 v4 @ 2.40GHz" CPU. It took approximately 23.5 milliseconds (ms) to preprocess an EEG segment of two-second length. Following the preprocessing, our network required about 10.0 ms to process the resulting features. In total, the solution took about 33.5 ms to handle an input EEG signal segment arriving every second. This signifies that our algorithm can process incoming EEG data in real-time, allowing for continuous monitoring and rapid response. 
	
	However, considering the limitations of hardware capabilities in wearable devices, it is infeasible to run our algorithm directly on them. Instead, a more practical approach is to utilize a remote computation framework in which the wearable devices are primarily responsible for collecting the EEG signals and transmitting them to a remote server.
	
	\section{Limitations}
	In this section, we will discuss the limitations of our study. Following the seizure-independent division, the training data for some seizures becomes limited, while data for other seizures becomes abundant. It is challenging to assess how the imbalance in the training data may affect the training results. Therefore, the seizure-independent division may not be feasible for patients with limited and imbalanced seizure data, and ample EEG data is crucial for implementing seizure-independent division in clinical practice. Beside, we have not yet determined why the channel selection method failed for Patient 22 in the even division, but succeeded under the seizure-independent division.
	
	Additionally, the features and configurations we selected for prediction may not be the optimal solution. We have not exhausted all possible time and spectral features or their combinations, and we simply set the length of EEG signal segment according to the configuration of previous studies. After introducing channel selection, Patient 04 and Patient 07 had the lowest sensitivity of 33.3\%. This reduced sensitivity may be attributed to the limited feasibility of the selected features for these two patients.  
	
	Moreover, the number of patients is limited due to the lack of more publicly available datasets that include long-duration scalp EEG signals, so statistical analysis has not been conducted.
	
	\section{Conclusion}
	In this study, a novel two-stage channel-aware Set Transformer Network was proposed to conduct seizure prediction and channel selection concurrently. After implementing channel selection, a mean sensitivity of 80.1\% with 0.11 false alarms per hour was achieved in the even division. Additionally, the number of electrode channels required for prediction decreased to an average of 2.8, effectively increasing the feasibility and acceptance of predictive devices among epilepsy patients. A more rigorous seizure-independent division of EEG data, which can avoid the adjacency of training and test data, was conducted and evaluated, and the experimental results support our concerns about the traditional even division. In summary, our work in channel selection, seizure onset prediction, and data division methods contributed to epileptic seizure prediction. The code of this study has been released at ``https://github.com/RuifengZheng/Two-stage\_Set\_Transformer".
	\\
	
	\noindent
	\textbf{Declaration of Competing Interest}\\

	\noindent
	None.
	
	\bibliographystyle{IEEEtran}
	\bibliography{refs}
	
\end{document}